\documentclass[preprint,12pt,authoryear]{elsarticle}

\usepackage{amssymb}
\usepackage{amsmath}
\journal{Arxiv}

\begin{document}

\begin{frontmatter}



\title{On Optimality and Human Prediction of Event Duration in Real-Time, Real-World Contexts} %% Article title


\author{Mark G. Orr} %% Author name

\affiliation{organization={Florida Institute for Human and Machine Cognition},%Department and Organization
            addressline={40 S. Alcaniz St}, 
            city={Pensacola},
            postcode={32502}, 
            state={Florida},
            country={USA}}

%% Abstract
\begin{abstract}
The focus of the current work concerned the psychological processes that underlie prediction of an event’s duration.  The objective was to push forward existing psychological theory on event duration prediction, something made possible by the unique features of our data context.  The provisional findings suggested that the prior, existing theoretical mechanism of event duration prediction is incomplete because: (i) it doesn’t support adaptive responses when event duration judgments are dependent, (ii) it doesn’t  afford the integration of new, on-the-fly, information. Our findings suggest specific directions for future research.
\end{abstract}


%\item Insight into human predictions of event duration (prediction of how much longer a process or event will last).
%\item Phenomena is one that plays an important role in everyday life (planning) and extraordinary life (pandemic preparedness for governments).  
%\item Provides scientific understanding of predictions of event duration in real-time, naturalistic setting.
%\item Our analysis provides provisional evidence, using a mathematical model, that humans adapt to real-time context and dynamics.   

%% Keywords

\end{frontmatter}



\section{Introduction}
\label{sec1}
%% Labels are used to cross-reference an item using \ref command.
Human judgments of future events are ubiquitous and span the mundane (e.g., will the oncoming traffic light change from green to yellow) to the extraordinary--e.g., decisions made during military operations. The study of human forecasting, fittingly, draws upon myriad domains and literature. Our report focuses on what is called prediction of event duration: forecasting how much longer an event will continue in reference to how long it has lasted thus far.  Examples are:   "We've been at war for 2 months; how much longer will the war last?"  "A train has been traveling for 20 minutes; what will be the total time of the trip?"
 
Here, we explore event duration prediction in real-world epidemics via two rounds of an online forecasting tournament focused on the COVID-19 epidemic.  Participants were asked to judge the timing of the peak of a wave of the epidemic.  Measurement-wise, the data were impoverished, offering little beyond the actual predictions.  Thus, we adopted the rational analysis perspective \citep{anderson2013adaptive, Lieder2020, marr2010vision}, a first step towards understanding the psychological processes that drive behavior. Rational analysis puts forth an optimal computation (a rational model) in relation to a task and addresses the degree to which human performance matches the optimal computation (see \cite{Tauber2017} for detailed exposition).  

Rational models show promise in explaining event duration prediction:  (i) humans can estimate an event's duration optimally given little context and (ii) prior knowledge integrates with new, observed data in an optimal way \citep{GriffithsTenenbaum2006,GriffithsTenenbaum2011,Tauber2017}.  (These results come with some theoretical contention, e.g., \cite{MozerPashlerHomaei2008}).  The basis of these models is a Bayesian decision model derived from: 
\begin{equation}\label{eq:posteriorfori}
    P(t_{total} \vert t_{past}) = \frac{P(t_{past} \vert t_{total}) P(t_{total})} {P(t_{past})}
\end{equation}
where $t_{total}$ is a value of an event's \textit{total} duration, coming from a distribution of durations; $t_{past}$ is the duration at which a decision is made, or if post-hoc, the point for which the decision is made; $P(X)$ stands for probability of $X$.  We define the decision function as the expected value of the probability distribution generated from Equation \ref{eq:posteriorfori} over the distribution of all $t_{total}$. We call the output of this function $t_{predicted}$; i.e., this is the rational model's prediction of the duration of the event.  

Defining the Bayesian prediction function (borrowing from \cite{GriffithsTenenbaum2006}) as $t_{predicted}$ over all values of $t_{past}$,  shows concordance with human predictions under experimental conditions \citep{GriffithsTenenbaum2006} in which human predictions are provided in reference to a single event (e.g., how much longer will someone live if they are now 55 years old).  Further, experimental evidence supports the hypothesis that human event duration prediction is affected by both prior, durable knowledge and new observations; in terms of the the Bayesian decision model the new observations are integrated into the likelihood function $P(t_{past} \vert t_{total})$ \citep{GriffithsTenenbaum2011}.

We applied the Bayesian decision model to a real-time, real-world context.  In our data setting, participants were allowed to make repeated predictions over time, during a single event (the Omicron wave of COVID-19).  This observational setting--uncontrolled repeated observations in real-time--afforded a unique, applied context for understanding two key issues for the Bayesian decision model.  

First, the setting had the potentiality of eliciting $t_{predicted}$ as $t_{past}$ approached $t_{predicted}$, a feature which surfaces a paradoxical conclusion of the Bayesian decision model.  When an agent is invariant in its prediction over time ($t_{predicted}$), the expected value of the prior distribution must decrease as $t_{past}$ approaches $t_{predicted}$\footnote{We will provide details of this result in the Methods section.}.   This reduction in prior, what we call the \textit{prior crash}, is paradoxical not because priors must be fixed but because the change in prior would be self-organized, without information external to the model.  Rapid changes in the prior are typically assumed to be driven by changes in the environment (see \cite{Sohn2021,Prat2021} for examples).   Contrast this with another limiting case.  When the prior is fixed (instead of the prediction), the prediction $t_{predicted}$ increases as $t_{past}$ approaches it.  Taken together, the invariant prediction and the invariant prior are just flip-sides of the same model. The former forces the prior downward to maintain a constant prediction; the latter forces the prediction upward to maintain an invariant prior. 

The paradoxical feature of the Bayesian decision model is an issue only to the degree that it is observed.  As $t_{past}$ approaches $t_{predicted}$, humans may adapt their predictions, either in an effort to minimize the degree of change in the prior or some other motive (we leave the motive for the discussion). So, the first objective of our work was to \textit{observe how participants adapt their predictions in real-time and to understand the relation of the un-observable theoretic component, the prior, to the human predictions}.  

Second, forecasting tournament participants were provided with a url to the COVID-19 daily-case counts as a potential source of information.  Thus, the forecasting context afforded exploring the role of information external to the Bayesian decision model.
Thus, our second objective was \textit{to explore possible relations between external data and the human predictions}. If the human predictions were correlated in some manner with the epidemiological case-counts, then it would suggest re-consideration of the rational model described above, depending on the nature of the relation.  A priori, it is not obvious how to integrate such information into the Bayesian decision model we used.

%% Use \subsection commands to start a subsection.
%\subsection{Example Subsection}
%\label{subsec1}

%No Subsection here.

%% Use \subsubsection, \paragraph, \subparagraph commands to 
%% start 3rd, 4th and 5th level sections.
%% Refer following link for more details.
%% https://en.wikibooks.org/wiki/LaTeX/Document_Structure#Sectioning_commands

%% Use figure environment to create figures
%% Refer following link for more details.
%% https://en.wikibooks.org/wiki/LaTeX/Floats,_Figures_and_Captions
%\begin{figure}[t]%% placement specifier
%% Use \includegraphics command to insert graphic files. Place graphics files in 
%% working directory.
%\centering%% For centre alignment of image.
%\includegraphics{example-image-a}
%% Use \caption command for figure caption and label.
%\caption{Figure Caption}\label{fig1}
%% https://en.wikibooks.org/wiki/LaTeX/Importing_Graphics#Importing_external_graphics
%\end{figure}

\section{Methods}\label{methods}

We present our work in two studies.  Study 1 provides a descriptive overview of the human predictions during the course of three rounds in the forecasting tournament, both in aggregate and idiographically (for some individuals).  This will provide a sense of the data structure and the performance of the forecasters.  The primary focus for Study 1, however, was to tackle our first objective: observe how participants adapt their predictions in real-time and to understand the relation between the unobservable theoretic component, the prior, and the human predictions.  Study 2 addressed our second objective: to explore possible relations between external data, the human predictions and, potentially, to components of the rational model (e.g., the prior).  To this end, we used the same human predictions from Study 1 and provided an analysis of the relation of the human predictions to the epidemiological case-count in Virginia. Next, we provide general methodological details that apply to both studies prior to detailing the specific methods used in each study.

\subsection{Forecasting Interface}\label{interface}
Participants were not restricted in terms of the number of forecasts they could make, given they were within the forecasting window (the window of time for which forecasters were permitted to submit predictions).  The interface presented to participants the prediction window of the question (the window of time for which predictions were made) as a timeline; overlaying the timeline was a visual representation of a probability density distribution (logistic in form).  The participant entered a forecast by manipulation of three horizontal sliders, two of which adjusted the width (left, right) of the probability distribution and one of which adjusted its center.  Further, the participants could generate a mixture distribution for their prediction as a layered and weighted combination of up to five separate distributions.  
%MUST GET ANSWER ON NON-SYMMETRY OF HAND GENERATED FUNTION VS LOGISTIC.
%Formally, this is:
%\begin{equation}
%    f(x) = \sum_{k=1}^{5} \pi_k g(x\vert \mu_k,s_k)
%\end{equation}
%where $\pi$ represents weights for each of the five (up to five) distributions.  Each distribution was defined as:
%\begin{equation}
%    g(x\vert \mu_k,s_k) = \frac {e^{-x(x-\mu)/s}} {s(1+e^{-x(x-\mu)/s})^2}
%\end{equation}
%The \pi_i were constrained so that their sum was $\le$ 1.
In addition to the visual of the prediction distribution, the participant saw the median, and quartiles of the distribution both visually (as vertical lines in the probability distribution) and numerically.  The questions under investigation in this article were discrete in nature (day); the numerical value of the prediction was presented at the day resolution to the participant as a summary measure of the participants probability distribution.  

\subsection{Data Structure and Major Constructs}\label{data-structure}
These data were extracted from the first three rounds of an online prediction tournament run by Metaculus and using the Metaculus forecasters pool as the participants (we describe the pool in section Participants).  The dates available for forecasting (the forecasting window) per rounds 1-3 were, respectively: 11-12-2021 to 12-03-2021; 12-03-2021 to 12-24-2021; 12-24-2021 to 01-14-2022.  The prediction window (the horizon of the forecast) for the forecasts per rounds 1-3 were, respectively: 11-12-2021 to 02-03-2022; 12-03-2021 to 02-25-2022; 12-24-2022 to 03-18-2022.  Each forecast made by a participant was in answer to the following question:  "When will the peak .. occur ". For the remained of this article, we call this the duration of the epidemic as short-hand for the duration from start to the peak of the epidemic curve. 

To generate $t_{past}$ we defined $t_0$ to be the beginning of the Omicron wave of COVID-19 and set it to November 29, 2021; this date was based on a reasonable estimate considering the daily reported case-count curve in Virginia (see \cite{VDHOnline}).  To construct $t_{predicted}$ for each forecast, we first transformed the participant input to the forecasting interface (described above) into a single date value that approximated the observed median of a prediction.  The data structure provided by the forecasting interface was in the form of a discrete probability mass function of ordered time intervals over an 84 day prediction window.  This generated 101 evenly spaced time intervals of the length of 20 hours, 9 minutes and 36 seconds (this interval times 101 = 84 days) each with a probability.  We defined the median of the prediction to be the discrete value $X$ for which $X \le 0.50$ when transforming the discrete probability mass function into a cumulative probability mass function.  In short, this procedure defined a single value, the approximate median value, as predicted date. Then, to construct $t_{predicted}$, we subtracted $t_0$ from the predicted date (e.g., Jan 2, 2022 minus November 29, 2021 $= t_{predicted}$ of 34 days).  Further, another useful construct is the prediction horizon, what we call the \textit{human horizon}; we defined this to be the number of days between the date on which a prediction was generated and the predicted date of the prediction.  This measure was useful because it represents the distance between $t_{past}$ and $t_{predicted}$ directly, a measure that theoretically important in terms of the potential for the rational prediction model to generates seemingly paradoxical behavior, as described in the Introduction.  In particular, in our analysis we were concerned with behavior of the rational model when the horizon was small.  We also extracted the date the prediction was generated by the participant, a measure that did not require transformation.

In short, the forecasting tournament data provided the date the prediction was generated by the participant (e.g., December 20, 2021) and the predicted date of the epidemic peak (e.g., Jan 2, 2022). Given a pre-defined beginning of the epidemic curve, $t_0$, we could generate both $t_{past}$ and $t_{predicted}$ as required by our theoretical model.  Further, we use the human horizon to understand the relation between the human predictions and the rational model when the human horizon is small.

The other major data structure reflected a measurement of the daily-case count for COVID-19 in Virginia.  These publicly available data were provided by the Virginia Electronic Disease Surveillance System (VEDSS) \citep{VDHOnline} and are entered by 5:00 PM the prior day (but are subject to change as quality assurance was ongoing).  VEDSS adopted the CDC COVID-19 2021 Surveillance Case Definition \citep{CDCguide} on September 1, 2021.  The quality assurance methods for VEDSS in respect to these data can be accessed here \citep{VDHquality}.

The raw form of these data is the cumulative number of cases on each day.  We transformed this into the daily number of cases per day by computing daily differences.  We then computed the rolling seven-day average over these raw data.  Because we had data for days well beyond both end-points of interest (11-30-2021 and 01-14-2022), our data structure on those days reflects the rolling seven day average prior to and after those end-points where appropriate.

\subsection{Participants}\label{participants}
All participants were users on the Metaculus forecasting platform (see \citep{metaculus}).  Metaculus is an online platform that hosts a community of forecasters (N is approximately 15,000). It is publicly accessible (with free registration for an active account) and hosts individual questions and tournaments with monetary prizes.  The Metaculus platform does not collect demographic data on the forecasters.  Further, for the participants reported in this article, we could not access summary information on the distribution of competence from prior forecasts. 

\subsection{Pre-processing of Forecasts}
The design of the forecasting tournament necessitated some pre-processing of these data; we selected/filtered based on criteria applied to the predictions not to participants. We started with 471 predictions that were generated by 47 participants.  The first step was to filter out predictions that were using the end-points of a round's prediction window, what we dub end-point forecasts; left end-point forecasts were at the earliest point of the prediction window; right end-point forecasts were at the latest point.  Because the prediction windows were bounded, we could not know if a end-point forecast was for the endpoint or beyond it.  None of the predictions fit this criteria.  The second step was to filter out the predictions that were prior to our definition of the start of the epidemic wave, $t_0$.  This reduced the number of predictions to 403 and the number of participants to 39.  The final filtering process was to remove predictions that were less than $t_{past}$ because these are theoretically impossible (impossible by the theory).  This removed two predictions for a final usable total of 401 predictions generated by 39 participants.  The distribution of the number of forecasts per user was highly right-skewed (mean=10.33, std=15.98, median=3, skew=2.15).

\subsection{Key Features of the Theoretical Model}
Our theoretical model, described in the Introduction, was borrowed directly from \citep{GriffithsTenenbaum2006} and differed only in its implementation (we describe the implementation in the next section).  We describe two key features of the model here: (i) the paradoxical \textit{prior crash} that was outlined briefly in the Introduction and (ii) the theoretical integration of new observations into the decision process and how statistically independent and dependent observations effect the model differently.  

\subsubsection{Invariant Predictions vs Invariant Priors}
We introduced the somewhat paradoxical feature of the Bayesian decision model when $t_{past}$ approaches $t_{predicted}$ in the Introduction.  Here we provide a detailed description of how this operates (please refer to Equation \ref{eq:posteriorfori} for this discussion).    This is best illustrated via an hypothetical example.  A person, Pat, is traveling on a train on a familiar route. After 20 minutes from departure, Pat is asked by another passenger how long will the train trip last (i.e., what will be the total duration in minutes).  For this example, $t_{past}$ is 20 minutes and $t_{total}$ is one value from a distribution encoded in memory by Pat's past experience on this train route, e.g., 45 minutes. Pat's response, according to the Bayesian decision model, is formulated as the expected value of the probability distribution generated from Equation \ref{eq:posteriorfori} over the distribution of all $t_{total}$. We call this decision $t_{predicted}$ to capture the notion that the rational model predicts a value of $t$ to be the duration of the current train trip.  In Bayesian terms, $P(t_{total})$, over all instances of $t_{total}$, defines the prior distribution; $P(t_{past} \vert t_{total})$ is the likelihood; $P(t_{total} \vert t_{past})$, over all instances in the prior, is the posterior distribution for a single value of $P(t_{past})$.

\begin{figure}
    \centering
    \includegraphics[width=0.75\textwidth]{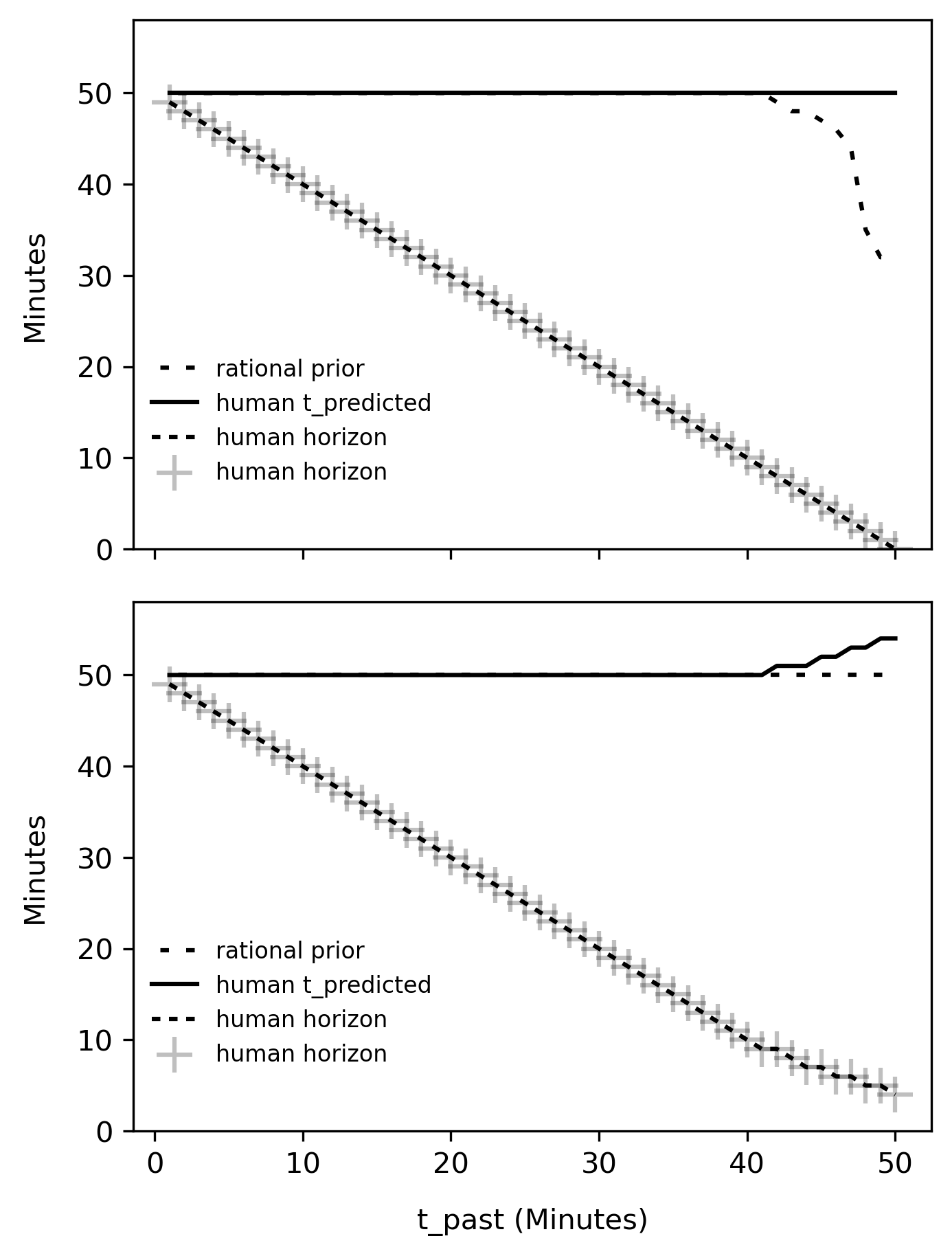}
    \caption{\footnotesize{The theoretical relation between rational priors, $t_{past}$ and $t_{predicted}$ given the rational decision model for a hypothetical human riding on a train.  The x-axis represent the time at which a prediction is made, $t_{past}$;  the black, solid line represents the human decision, $t_{predicted}$; the black, large dashed line shows the prior, given $t_{past}$ and $t_{predicted}$, generated by the rational decision model; the human horizon reflects the amount of time until the predicted event from $t_{past}$ (i.e., its how far forward is the prediction from the moment it is given).  The top panel illustrates the invariant human decision case (a fixed $t_{predicted}$); as $t_{past}$ approaches $t_{predicted}$, the prior necessarily drops (what we call the "prior crash").  The lower panel shows the invariant prior case (a fixed prior distribution); as $t_{past}$ approaches $t_{predicted}$, $t_{predicted}$ increases.  
    }}
    \label{fig:PriorShift2Panel}
\end{figure}

\begin{figure}
    \centering
    \includegraphics[width=0.75\textwidth]{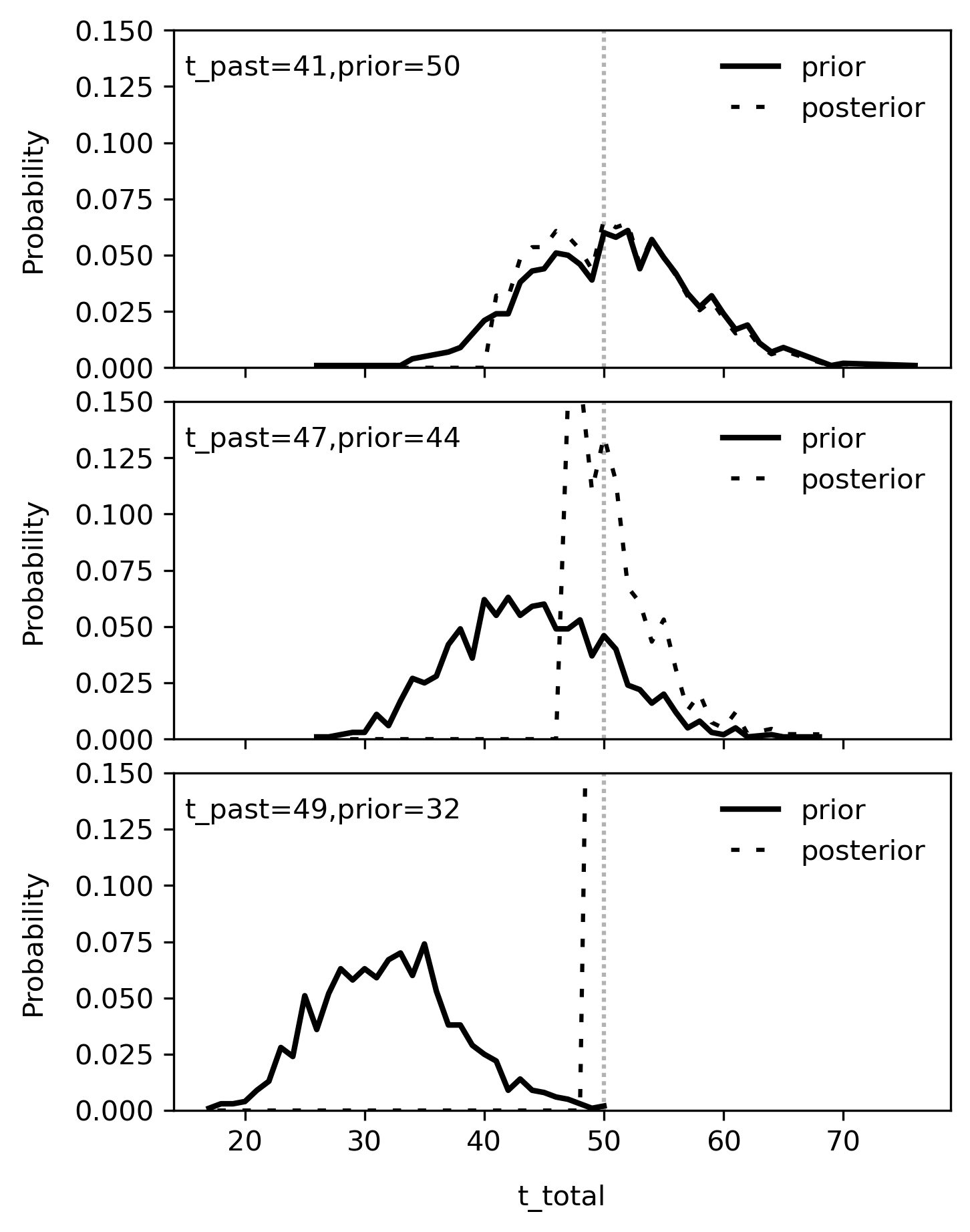}
    \caption{The relation between prior and posterior distributions given three instances of the rational decision model; these instances were derived from the scenario shown in the top panel of Figure \ref{fig:PriorShift2Panel} for the condition of an invariant human decision, $t_{predicted}$, while riding on a train.  In each panel, the black solid line shows the posterior distribution; the black large dashed line shows the posterior distribution; the vertical grey small dashed line annotates the fixed invariant decision of 50 minutes.  Moving from the top to the bottom panel, this figure shows the leftward shift in the prior as $t_{past}$ increases from 41 to 47 to 49 minutes.  During this shift the median of the prior distribution shifts from 50 to 44 to 32 minutes while the median of the posterior distribution is maintained at 50 minutes.
    }
    \label{fig:PriorShift1}
\end{figure}

Let us continue on our journey with Pat.  Imagine that Pat is riding on a train and is probed to provide a prediction of the duration of the trip (i.e., $t_{predicted}$) at at a frequency of once per minute.  Further, assume that Pat generates an invariant value for $t_{predicted}$ of 50 minutes whenever probed.  In this scenario, a conclusion of the rational model is that Pat's prior must decrease as $t_{past}$ approaches $t_{predicted}$. This scenario is illustrated in the top panel of Figure \ref{fig:PriorShift2Panel}.  Put simply, by the rational prediction model, Pat's prior must decrease rapidly when Pat's prediction is invariant.  Figure \ref{fig:PriorShift1} shows the time evolution of the prior under this scenario, something that is recoverable if we know both $t_{past}$ and $t_{predicted}$. What we see here is that to maintain an invariant $t_{predicted}$ as $t_{past}$ approaches it, the prior distribution shifts rapidly to the left.   

Another side of this paradoxical feature of the rational model is the limiting case when the prior is fixed (shown in the lower panel of Figure \ref{fig:PriorShift2Panel}), the prediction $t_{predicted}$ increases as $t_{past}$ approaches it.  For this case, we imagine that Pat has an invariant prior the effect of which is that Pat's prediction of when the train trip will end, $t_{predicted}$, increases as $t_{past}$ approaches it.  This is similar to asking Pat a different but common question. Imagine we ask Pat to predict a person's life-span; as $t_{past}$ (the current age of a person) approaches the expected value of the prior, the predicted life span increases--e.g., given that someone is 79 years of age (the expected value in the US \cite{Arias2019}), Pat's prediction, given the rational decision model, is a bit greater than 79.

%%%%

\subsubsection{The Integration of New Observations}
The integration of new observations, in terms of the decision model, operates on the likelihood function defined as:
\begin{equation}\label{eq:likelihoodN}
   \left(\frac{1}{t_{total}} \right)^n
\end{equation}
(when $t_{total} \ge t_{past}$) where $n$ is the number of observations (see \cite{GriffithsTenenbaum2011} p. 729 for derivation).  Thus, as the number of observations increases, the predicted $t_{predicted}$ decreases, assuming the prior is invariant.  This is the assumed case for statistically independent observations.  However, the experimental evidence suggested that for dependent observations, $n$ is effectively equal to one, suggesting provisionally that humans are sensitive to the generating process of new observations \citep{GriffithsTenenbaum2011}.  In short, the theoretical model and associated experimental results suggest that only independent observations can effect change in the likelihood. 

We raise this issue here because our observational context yielded repeated forecasts for individuals that are statistically dependent.  In reference to our hypothetical example, this is akin to asking Pat (from the example above) to predict the length of a single train trip repeatedly over the course of the trip at different values of $t_{past}$.

\subsection{Methods: Study 1}
Our methodology was descriptive but in a manner that included the theoretical unobservable constructs of the Bayesian decision model.  We observed participants predictions from the forecasting tournament in real-time, both in aggregate and as individuals, and recovered the priors implied by the rational decision model.  This afforded a view of the dynamics of human predictions and their respective priors over the course of the tournament.  In particular, we observed conditions such that $t_{past}$ approached $t_{predicted}$ and thus served our first objective well.  We describe next the method for recovering the prior from the participant forecasts. 

\subsubsection{Recovering the Prior from Observations}
The primary use of the theoretical model in Study 1 was to recover the best prior given a participant's prediction, $t_{predicted}$, and the value of $t_{past}$ at the time of the participant's prediction.  Our theoretical model, described in the Introduction, was borrowed directly from \citep{GriffithsTenenbaum2006} and differed only in its implementation. We used a discrete, sampled prior and numerically computed the prior probability mass distribution from it; the likelihood was also computed from the same probability mass distribution.  Specifically, we used the Poisson distribution for the priors with the following probability mass function: 
\begin{equation}\label{eq:poisson}
   P(X=k) =  \frac {\lambda^k e^-{\lambda}}{k!.}
\end{equation}
where $k$ was the number of days and $\lambda$ was the expected number of days in an epidemic wave.  For each value of $\lambda$, across the range from 20 to 200 days, we generated 1000 samples numerically, using the scientific software 'Numpy'\footnote{We decided to use numerical samples in place of exact results to allow for a simple method for integrating sample biases into future modeling efforts or for integrating empirical estimates of real infectious disease wave durations.} from which we could compute the sampled probability mass function over the values of $k$ that were sampled numerically for a given $\lambda$.  This had the effect of restricting the prior and likelihood to the sampled values, but we desired a simple, approximate value for this study.

To find the prior distribution that was the best candidate for generating the human prediction $t_{predicted}$, we generated a three-way table that represented all combinations of prior distributions (medians of), $t_{past}$, and the median of the Bayesian decision model output $t_{predicted}$\footnote{Note that $t_{predicted}$ can refer to the human decision or the output of Bayesian decision model; this term will be used for both interchangeably} for the values constrained by the expected values $\lambda$ of our priors (20 to 200 days).  To recover the best candidate prior for a human prediction, we searched the three-way table for the closest match--i.e., we found the prior distribution that would have generated $t_{predicted}$ given $t_{past}$.  For purposes of analysis, we reported the expected value of the prior's distribution as the measure for analysis. Henceforth, we call this measure the prior; the context of our use of the term prior should help the reader to disambiguate the meaning of the expected value of the distribution from the distribution itself.  

By design, this method does not restrict a single prior to an individual over the course of the tournament, but only to a single prediction within the tournament; we wanted to see the extent to which an individual's prior varied over the course of the tournament, if at all.  

\subsection{Methods: Study 2}
Further, to serve our second objective, we provide an analysis of the relation between the human predictions and the Virginia daily-case count data of the COVID-19 Omicron wave.  This analysis relied on the rational decision model in the sense that it was the theoretical model that suggested to us a useful measure to extract from the tournament data: prediction in the form of $t_{predicted}$.  

This study used two data structures: (i) all 401 human predictions in terms of the predicted duration of the epidemic, $t_{predicted}$; (ii) the daily-case count for COVID-19 in Virginia, henceforth called case-count data.  

The human predictions were averaged by day (all predictions in a 24 hour period were averaged together) and then fitted using linear polynomial regression (where datetime was transformed to a linear variable as the predictor; we do not provide the coefficients here, but will report them upon request.  The purpose was to provide a relatively low-dimensional and smooth representation of the temporal signature of these data for our next processing step. Using the polynomial solution, we then computed, numerically, its first and second differentials.  We used the same method for the case-count data. 

In addition, we used a change-point method for the case-count data, based on the sequential discounting auto-regression learning (SDAR) algorithm \citep{Yamanishi2002}, to detect change points in the case-count data as we transformed it, prior to the polynomial fit, using a discount rate of 0.01, order of 3 and a smoothing factor of 5 days.

\section{Results: Study 1}\label{results_s1}
The panels of Figure \ref{fig:PriorShift2Panel} provide some bounds on the range of what we expect to see in the results. It may be that the boundary cases are not found; in particular, if the prior crash, as we call it, is not observed, e.g., due to adaptations in the human predictions, then the paradox implied by the rational model might not be an issue in practical terms. 

\begin{figure}
    \centering
    \includegraphics[width=\linewidth]{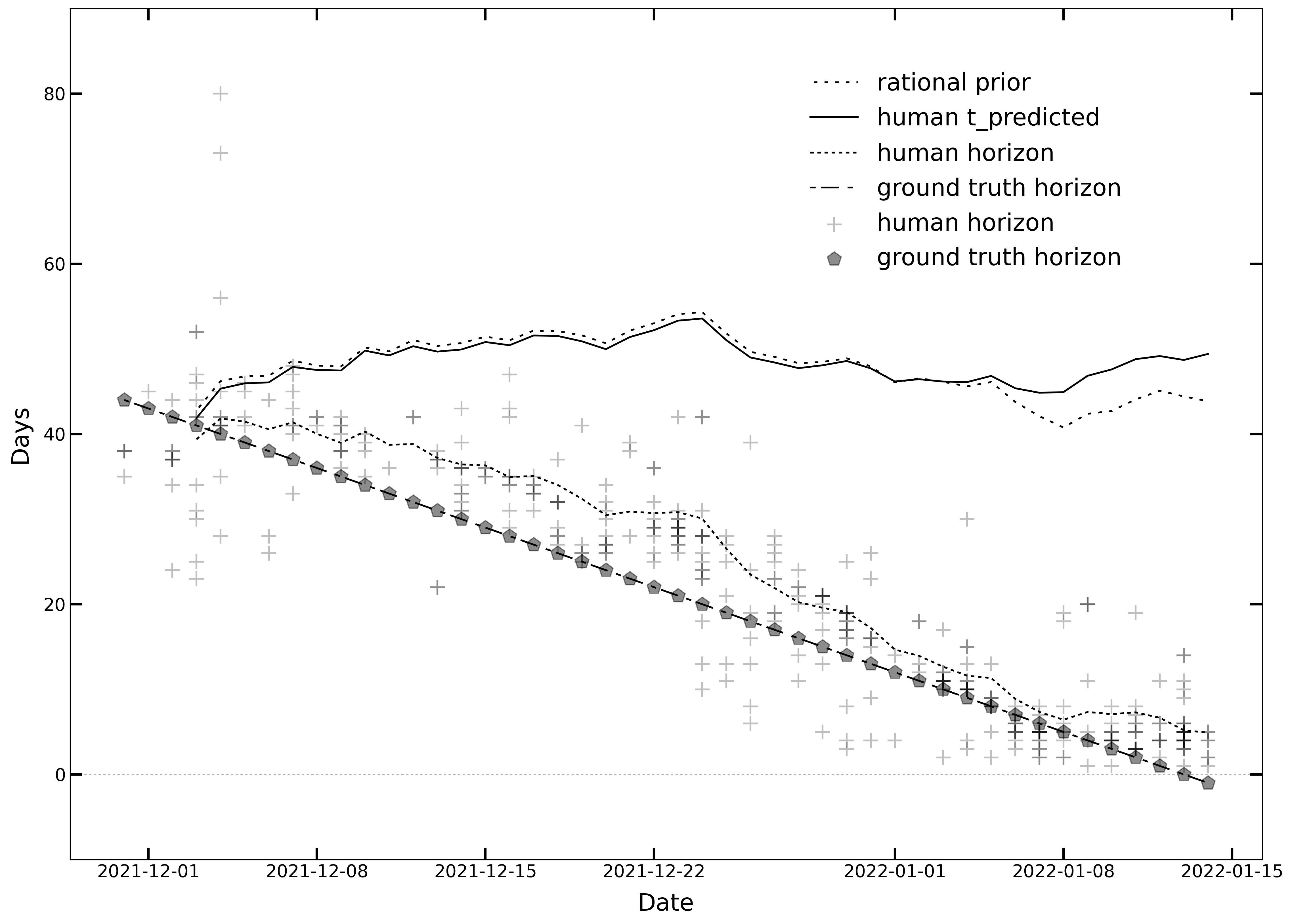}
    \caption{The two primary measures, human $t_{predicted}$ and the rational prior (using the median of the recovered prior), over the course of the tournament. (The y-axis shows days; the x-axis shows date of human prediction.)  The human horizon reflects human $t_{predicted}$ minus the date the prediction was generated by the participant.  The ground truth horizon shows the number of days until the actual epidemic peak in Virginia (January 13th, 2022) from each date in the figure; it is the ground truth equivalent of the human horizon.  All trend lines (except the groud truth horizon) were computed as 4-day moving averages. 
    }
    \label{fig:All_Ss_t_tot}
\end{figure}

The three key measures for this study were the human predictions $t_{predicted}$, the human horizon and the theoretical prior (as described in the Methods section, we are reporting the expected value of the recovered prior distribution).  Figure \ref{fig:All_Ss_t_tot} shows the aggregated result for Study 1.  We present these aggregated results in part to orient the reader to the nature of these data and the method of our descriptive analysis.  Keep in mind the following when referring to Figure \ref{fig:All_Ss_t_tot}:  (i) we use the date of prediction, which is constrained to be within the forecasting window, not the prediction window for the x-axis (see Methods for details of these measures); (ii) given (i), $t_{predicted}$ is the predicted total duration of the epidemic for a specific date, (iii) ground truth horizon was provided as a point of reference; it reflects what would be an accurate horizon given the actual epidemic peak of January 13th, 2022 and is directly comparable the human horizon in its meaning.  

The general features of this figure are as follows: (i) the human predictions of the duration of the epidemic, $t_{predicted}$, systematically increase to about half-way through the forecasting window, followed by a decline; the final portion of which shows an increase; (ii) most of the forecasting period shows a human horizon somewhat greater than the ground truth horizon, except for the beginning of the forecasting window and somewhat near the end of the forecasting window(around January 7th, 2022); (iii) the human horizon shows an upward trend during the end of the forecasting window.

In terms of the objective of Study 1--to understand the relation between $t_{predicted}$ and the prior, with emphasis on predictions that have a small horizon--we see in Figure \ref{fig:All_Ss_t_tot} some signature of the predicted \textit{prior crash}, but in attenuated form.  In short, the observations support, provisionally, some middle ground between the two limiting cases that were described in the Introduction (the constant prior vs the constant decision).  The upward trend of the human horizon toward the end of the forecasting window is suggestive of an adaptation on the part of the forecasters as the horizon closes.  

\begin{figure}
    \centering
    \includegraphics[width=\linewidth]{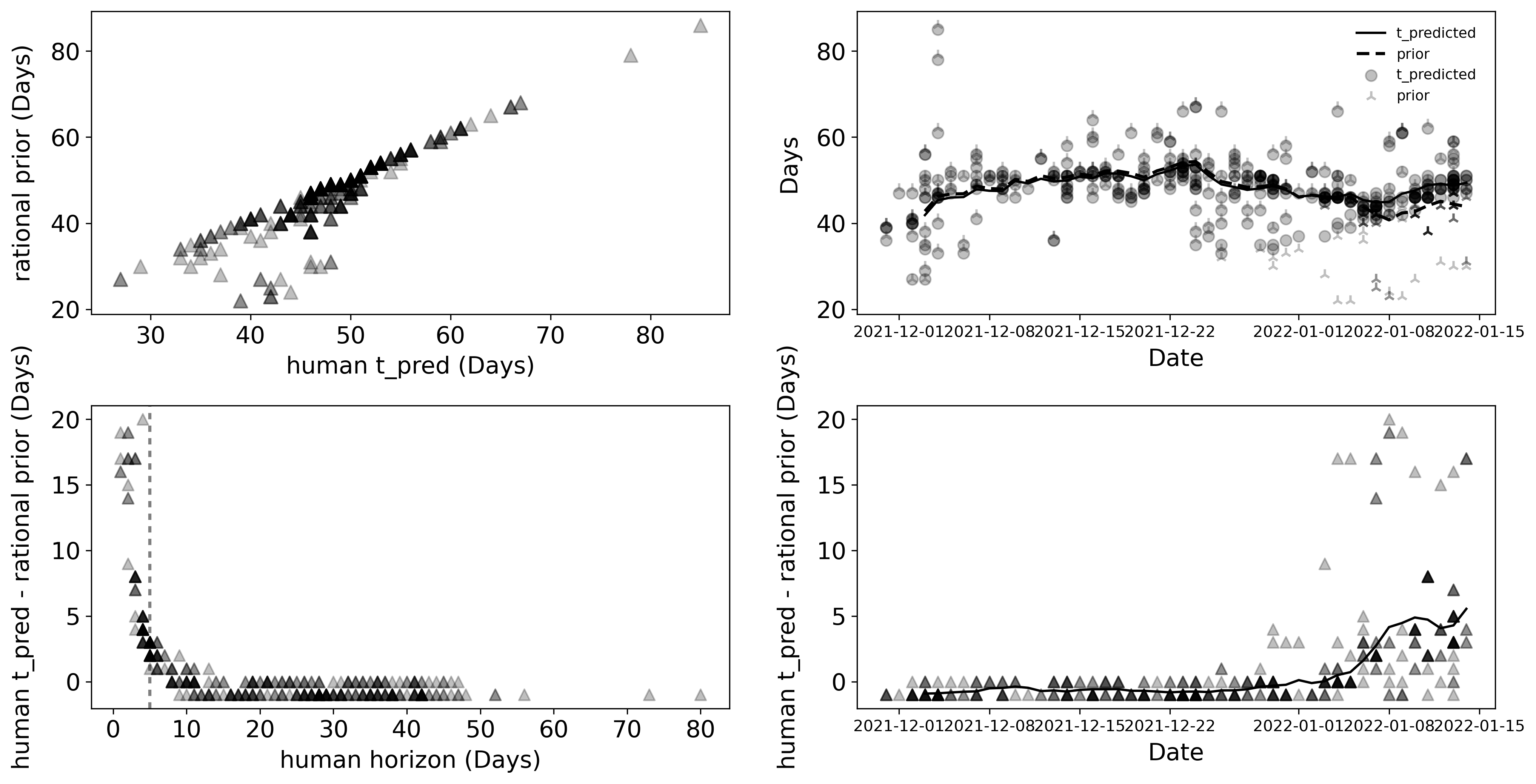}
    \caption{The top-left panel shows the relation between the human predictions and the rational prior; The top-right panel shows this relation over time. The lower-left panel shows the difference between the human prediction and the prior as a function of the human horizon; the lower-right panel shows this result over the forecasting window.}
    \label{fig:All_S_Scatters}
\end{figure}

This provisional conclusion is better understood when considering the non-aggregated data (showing all 401 predictions) as shown in Figure \ref{fig:All_S_Scatters}.  In the top-left panel of this figure, we see that the relation between the human predictions and the rational prior are close in value for many of the predictions with a subset of predictions that show larger values for $t_{predicted}$ compared to the prior.  The top-right panel of Figure \ref{fig:All_S_Scatters} shows this result over time, a plot that disaggregates the aggregate figure (Figure \ref{fig:All_Ss_t_tot}) and shows that the difference between a prediction and its prior only manifests near the end of the forecasting window.

The lower two panels in Figure \ref{fig:All_S_Scatters} are the most telling, in terms of the theoretical predictions.  The lower-left figure shows the difference between the human prediction and the prior as a function of the human horizon; it clearly shows that any substantial difference between the two only manifests with human horizons less than 5 days; the lower-right panel shows this result over the forecasting window.  

In summary, Figures \ref{fig:All_Ss_t_tot} and \ref{fig:All_S_Scatters} are suggestive of some adaptation on the part of the participants when the human horizon is small, a suggestion that is more strongly supported by the individual-level data from the most active participants.  We review these data next.

\begin{figure}
    \centering
    \includegraphics[width=\linewidth]{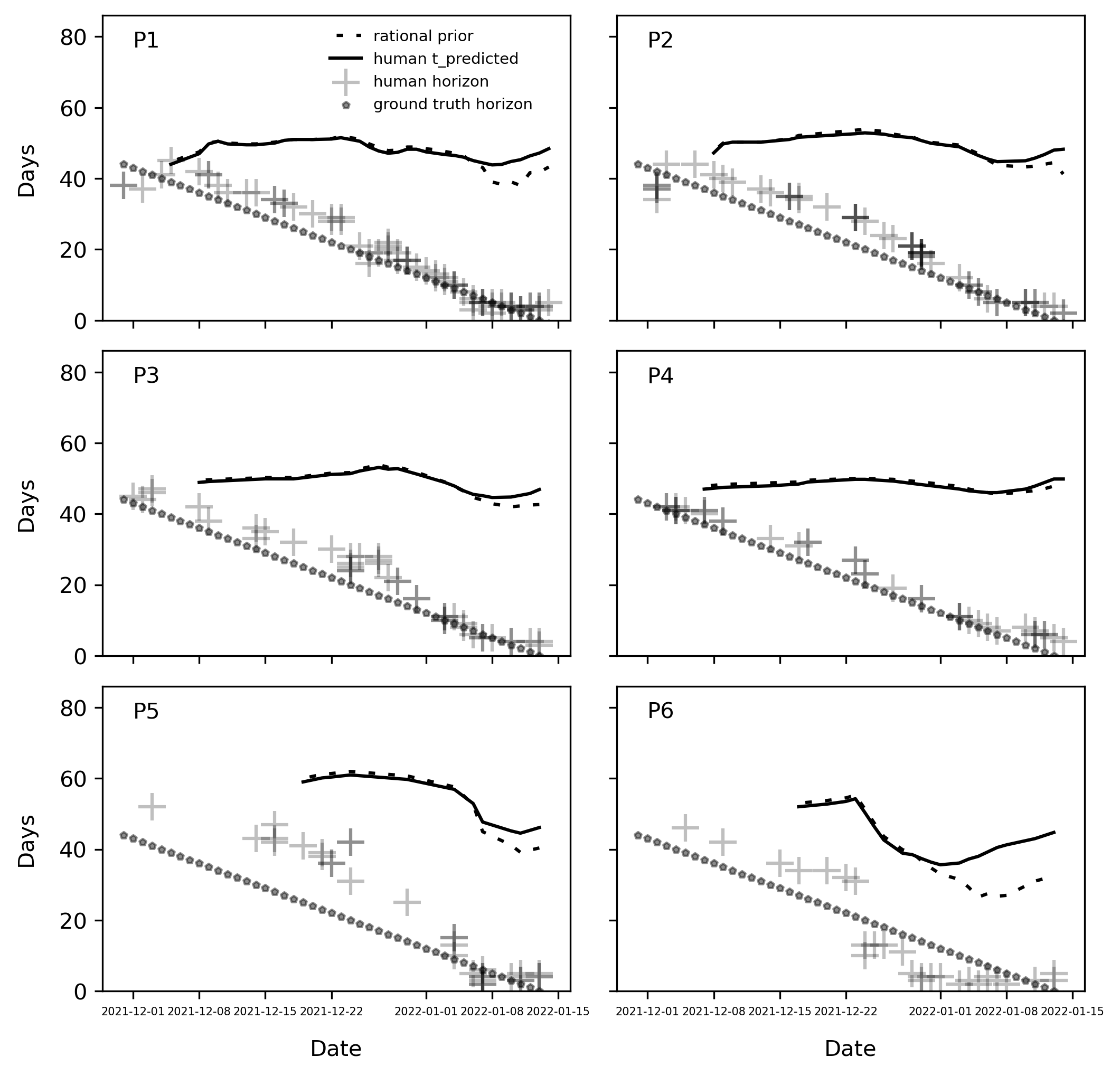}
    \caption{
    This figure depicts the same three components as in Figure \ref{fig:All_Ss_t_tot} for the six subjects with the highest response rate ( $> 15$ responses). Participant number is provided in the top-left corner of each panel.
    }
    \label{fig:Single_Ss_t_tot}
\end{figure}

\begin{figure}
    \centering
    \includegraphics[width=\linewidth]{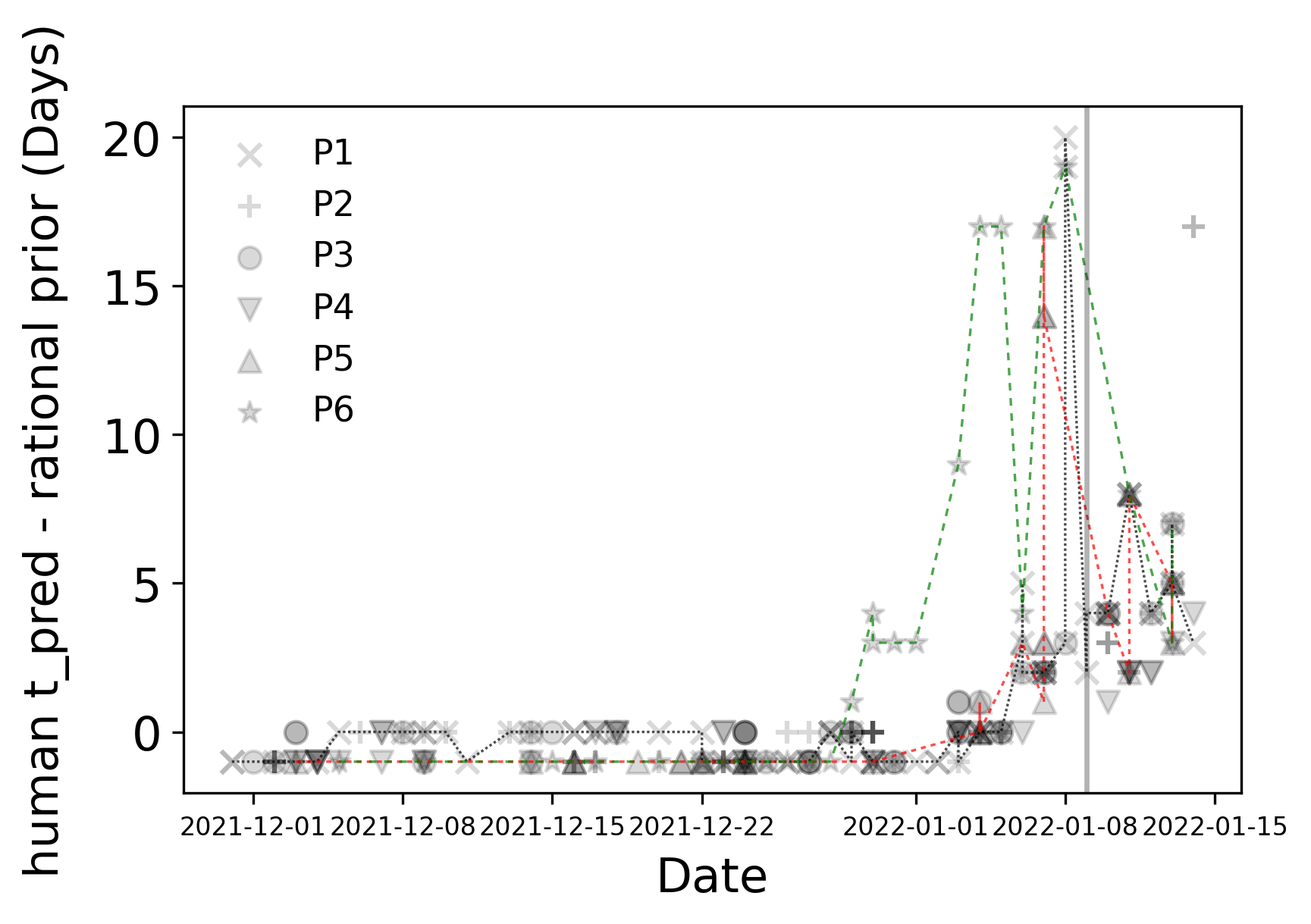}
    \caption{This figure is identical in form to the lower-right panel in \ref{fig:All_S_Scatters}; it shows the six most active participants, each labeled in the figure legend ("P" stands for participant).  The colored, dashed-lines show individual participants (green, participant six; red, participant five; black, participant one). The solid, vertical grey line falls on January 9th, 2022, the date at which (on and after) the predictions for participants one, five and six were post-peak.  See the text for details.}
    \label{fig:Single_Ss_Summary_2}
\end{figure}

Figure \ref{fig:Single_Ss_t_tot} shows a parallel result to Figure \ref{fig:All_Ss_t_tot} but for the most active subjects (those with more than 15 predictions).  All of the most active participants show some separation of the prior from the human predicted duration of the epidemic, something that is illustrated to a larger degree for participants numbers one, three, five and six.  Further, all of these participants (excepting participant number 4) show the following two signatures of adaptation: an upturn of the predicted duration of the epidemic accompanied by a flattening of the human horizon.  Further analysis of the disaggregated data is shown in Figure \ref{fig:Single_Ss_Summary_2}, a replicate of the lower-right panel of Figure \ref{fig:All_S_Scatters}. For this figure, we added lines that represent the time course for three of the six participants to illustrate the temporal peak of the largest differences between the predicted duration of the epidemic and the prior.  There was a clear peak prior to the end of the prediction window (these are all small horizon predictions, as gathered in the lower-left panel of Figure \ref{fig:All_S_Scatters}) followed by a set of predictions that had a reduced difference between the predicted duration and the prior, likely driven by an increase in the value of $t_{predicted}$.  It is important to note that for each of these three participants, there were multiple predictions both falling within the peak window and post-peak. 

The only participant with large differences between the predicted duration of the epidemic and the prior whom did not show similar behavior was number two; however, this amounts to only only one prediction that peaked for this participant which came very late in the forecasting window.  It is plausible that participant number two would have exhibited a similar signature as the other three participants given time.  The remaining two participants (numbers three and four)  did not exhibit such large differences between their predictions and their recovered priors. 

\section{Discussion: Study 1}\label{discussion_s1}
The primary purpose of Study 1 was to understand, in a naturalistic, observational setting, the relation between human predictions of duration and the unobserved prior, given the Bayesian decision model we described in the Introduction.  A principal value of the observational context was that $t_{predicted}$ and $t_{past}$ were not mutually constraining.  We hoped to and did observe cases for which $t_{past}$ approached $t_{predicted}$--the interesting case in terms of the Bayesian prediction model and its relation to human predictions. 

We came to this study without precise predictions, but with bounds on what we expected to see as presented in Figure \ref{fig:PriorShift2Panel}: the invariant prediction and the invariant prior.  Neither was observed.  Instead, our observations were consistent with the notion that the participants were adapting their predictions as the difference between the predicted duration of the epidemic and the prior deviated.  We saw signatures of this in both the aggregate time-series and in the individual-level time series.  The most suggestive result was that presented in Figure \ref{fig:Single_Ss_Summary_2} where we see a clear reduction in the difference between predicted duration and the prior after they deviated a large degree. 

Study 1 presupposes that the Bayesian prediction model operates without knowledge of the difference between $t_{predicted}$ and the expected value of the prior; this kind of comparison is not computed in the model.  So, by claiming that we have provisional evidence for human adaptation, we are also claiming, strictly speaking, that the decision model computes something in relation to this difference.  We leave this for future experimental study in which proper control would be instrumented.

One final comment for discussion.  It is paramount to understand that there is no necessity for the prior to split from the predicted value; the participant can always increase their predicted duration to a value large enough so that the expected value of the prior matches it.  In short, pulling up on the predicted value always works, if you pull hard enough.  We did not see this in these data, but a less severe adaptation.

\section{Results: Study 2}\label{results_s2}

\begin{figure}
    \centering
    \includegraphics[width=\linewidth]{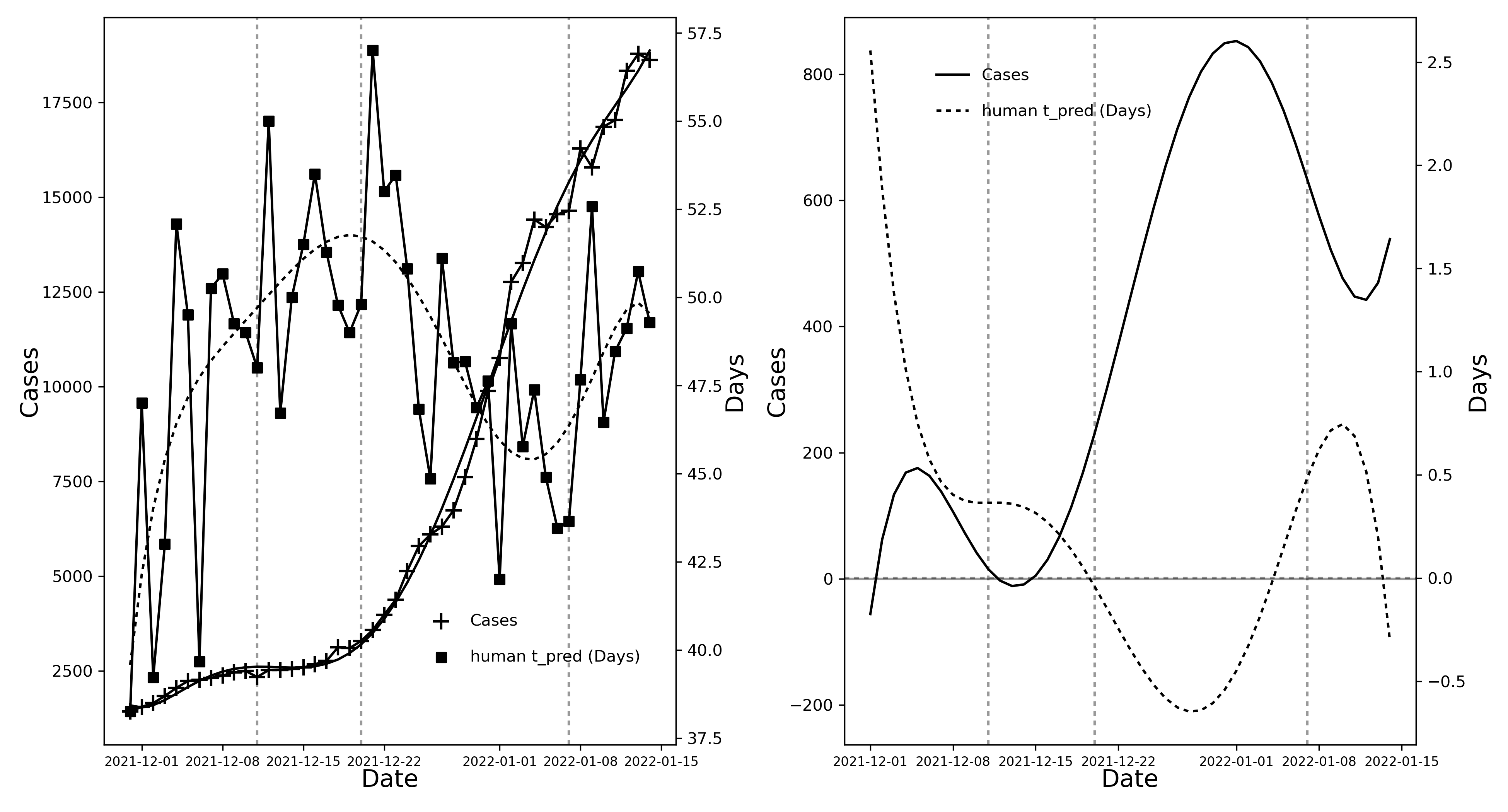}
    \caption{The left panel illustrates the case-count data and the human predictions super-imposed, along with the polynomial fits.  The three vertical lines indicate the change points in the case-count data.  The right panel shows the first derivatives of the polynomials for the case-count and human predicted durations.}
    \label{fig:Study_2_Multi}
\end{figure}

The key measures for Study 2 were the case-count data (after our transformation) and the set of human predictions of the duration of the epidemic ($t_{predicted}$).  Figure \ref{fig:Study_2_Multi} shows the results.  In the left panel, we see the case-count data and the human predictions super-imposed, along with the polynomial fits.  The three vertical lines indicate the change points in the case-count data.  The first noticeable feature in this panel is the large region of time after the second change point for which the case-counts and the human prediction move in opposing directions.  Another interesting feature is that the case-count curve was much flatter prior to the second change point compared to after it.  The final point of interest in this panel the similar slope of the two time-series in terms shortly prior to and after the third change point.  The second panel, showing the first derivatives of the case-count polynomials, provides a more nuanced understanding of the relation between the human predictions and the case-counts.  The major features of this panel are as follows:  (i) the strongest and most sustained growth of the case-counts is the period for which we see the strongest and most sustained negative growth of the human predicted durations; (ii) during the period mentioned in (i) we don't see positive growth in the human prediction until the case-count growth starts to slow.  In short, from the derivatives, we see two clear relations.  First, when the case-count growth is strong, the human growth is negative.  Second, when the case-count growth is slow or lessening from a high growth period, the human predictions show growth. 

\section{Discussion: Study 2}\label{discussion_s2}
The purpose of Study 2 was to explore the relation between external data and the human predictions of the duration of the event.  We did not have any predictions coming into this study in large part because the Bayesian decision model, as we specified it, does not afford the integration of data of this kind (or of any external information). Our discussion, therefore, will begin without consideration of this model. 

The relation between the human predictions and the case-counts, by some measure, was rather stark.  The human predictions seemed to respond to the growth of the case-counts in a way that was suggestive of how infectious disease operates.  In the beginning of an epidemic curve, slow growth represents a good deal of uncertainty: the epidemic might not take-off at all or it may be delayed significantly. Once an infectious disease agent shows strong growth, this can signal the potential for the infectious agent to rapidly burn through the susceptibles in the population, thus peaking sooner than expected.  Further, in a period of fast growth, if the growth again slows, it may signify a delay in the process.  It is plausible that humans are sensitive to this notion when making predictions about the duration of the epidemic.

\section{General Discussion}\label{discussion}
%%%PRACTICAL THEORY
\noindent
We open with a quote \citep{Lewin1943}:

"... there is nothing as practical as a good theory." (p. 118).

The rational theory proved practical, in the scientific sense.  Given impoverished naturalistic observational data, the theory afforded a mathematical decision modeling framework grounded in prior experimental work \citep{GriffithsTenenbaum2006,GriffithsTenenbaum2011}, a set of intriguing (if not paradoxical) predictions (Study 1) and a potential limitation (Study 2).  

The application of the model to real-world data returned the favor. First, it offers testable hypotheses.  In Study 1, the results suggested, in a highly provisional way, that participants may adapt to the difference between the $t_{predicted}$ and the prior.  This amounts to a testable, first-order hypothesis: are people actually sensitive to this difference?  To speculate, it should be possible to develop experimental protocols to measure this hypothetical phenomenon via existing experimental procedures and measures, e.g., \citep{sussman2007role}.  In Study 2, we saw a striking dynamic relation between the human predictions and the case-counts, one that is suggestive of an understanding of the epidemiological process on the part of the participants. 

Given said hypotheses, the model provides a second service--pointers for the construction of new theoretical processes.  The scope of this article does not warrant proposing a new model to accommodate external information (Study 2) or to incorporate the difference between $t_{predicted}$ and the prior (Study 1).  However, the flexibility of the theoretical approach is general as demonstrated by accommodation of a variety of findings across particular domains, e.g., anchoring bias \citep{Lieder2018}. 

The rational theory also proved practical, in the practical sense.  This work represents a step towards improving the psychological underpinnings of the kinds of human judgements that are directly relevant for a variety of policy, administrative or operational needs\citep{Galesic2021}.  The tournament from which we gathered our data was developed in conjunction with the Virginia Department of Health (VDH) to provide insights into future case numbers, vaccination rates, booster uptake and other key indicators for VDH COVID-19 operations.  Our work represents a slice of the tournament, but it begins the process of providing a scientific evidence-base for the forecasts, something valued in near real-time emergency decision-making \citep{Galea2021}.  In this policy space, improved human-machine collaborations may yield benefit.  

Another relevant policy application is towards developing high-fidelity population models of infectious disease.  Our work could provide the basis for the decision logic used by synthetic agents in large, at-scale agent-based models of infectious disease dynamics, an effort for which our laboratory is highly-experienced (University of Virginia).  Recent work has called for the integration of agents grounded in first-principles of cognitive science and psychology \citep{orrAttPolar2021,afrasiabi2019evaluating,orr2019multi,bhattacharya2019matrix,orrCovid19}.

In sum, the benefit of intermingling psychological theory with naturalistic data is clear from our work.  More varieties of this kind will benefit both psychological science and society.  
\section{Research Transparency}
 \subsection{General Disclosures}
Conflicts of interest: No authors declare conflicts of interest. Funding: Virginia Department of Health (COVID-19 Modeling Program VDH-21-501-0135). Artificial intelligence: No artificial intelligence assisted technologies were used in this research or the creation of this article. Ethics: This research received approval from a local ethics board (ID: 5657).

\subsection{Study One}
\subsubsection{Preregistration} No aspects of the study were preregistered.  
\subsubsection{Materials} The materials are not publicly available. We describe the materials in this article, section “Forecasting Interface.”  To gain an interactive sense of the materials, it would be useful to play with one of the existing Metaculus platform questions that has a similar format to the question used in this article (see {https://www.metaculus.com/questions/21209/dates-publicly-downloadable-large-ai-models} for an example).  
\subsubsection{Data and Analysis Code} Per a formal data use agreement (DUA) these data shall not be disclosed or released or otherwise granted access to any third party without the prior written consent of data provider (Metaculus).  Please contact the corresponding author for details about obtaining these data.   All analysis scripts are publicly available ({https://github.com/mark-orr/Cognitive\textunderscore{}Forecasting}).

\subsection{Study Two}
All aspects of Study One hold true for Study 2 in respect to the research transparency statement except the following.  The epidemiological data from Virginia that was used in Study Two is available for direct download here: {https://github.com/mark-orr/Cognitive\textunderscore{}Forecasting/blob/main/simulations/InputData/VDH-COVID-19-PublicUseDataset-Cases.csv}.

\bibliographystyle{elsarticle-harv} 
\bibliography{cas-refs.bib}



%\begin{thebibliography}{00}


%\bibitem[Lamport(1994)]{lamport94}
%  \textit{\LaTeX: a document preparation system},
%  Addison Wesley, Massachusetts,

%\end{thebibliography}
\end{document}